\documentclass[prl,twocolumn,floatfix,longbibliography,superscriptaddress,amsmath,amssymb]{revtex4-2} 
\usepackage{graphicx} 
\usepackage{bm}
\usepackage{bbm}
\usepackage{color}
\usepackage{amsmath}
\usepackage{amssymb}
\usepackage{color}
\usepackage{footnote}
\usepackage{comment}
\usepackage{hyperref}
\usepackage{subfigure}
\usepackage{appendix}
\usepackage[applemac]{inputenc}

\begin{document}
\title{Breathing mode inducing   dynamical   pairing in Kagome materials}

\author{Debmalya Chakraborty}
\email[]{debmalya.chakraborty@phy.iitkgp.ac.in}
\affiliation{Department of Physics, Indian Institute of Technology, Kharagpur, West Bengal 721302, India}

\author{Anushree Datta}
\email[]{anushree.datta@aalto.fi}
\affiliation{Department of Applied Physics, Aalto University School of Science, FI-00076 Aalto, Finland}

\author{Jorge Cayao}
\email[]{jorge.cayao@physics.uu.se}
\affiliation{Department of Physics and Astronomy, Uppsala University, Box 516, S-751 20 Uppsala, Sweden}

\date{\today} 
\begin{abstract}
The breathing mode in Kagome materials is a structural modulation that breaks inversion symmetry and has been shown to be a crucial source for intriguing phases in the normal state. In this work, we carry out a full classification of superconducting symmetries in kagome superconductors and demonstrate the emergence of odd-frequency dynamical Cooper pairs entirely driven by the breathing mode. We then show that odd-frequency spin-singlet Cooper pairs can be realized by controlling the breathing mode in kagome lattices with conventional spin-singlet $s$-wave superconductivity. Since odd-frequency pairing is intrinsically nonlocal in time, our results put forward the breathing mode for designing dynamical Cooper pairs in kagome materials.
\end{abstract}
\maketitle

Kagome materials have generated great interest in the past ten years due to their exotic electronic band structure featuring flat bands and Dirac cones, which, accompanied by an inherent frustration and strong correlations, harbor  exotic  ground states \cite{yin2022topological,wang2023quantum,1g9n-wm38,wang2024topological,wang2025intriguing}. Among the most salient examples, we find   charge density waves, topological phases \cite{yin2019negative,wang2024topological,wang2025intriguing}, and unconventional superconductivity \cite{hossain2025unconventional}. The characteristic of kagome materials is their two-dimensional lattice, which consists of a network of corner sharing triangles surrounding a central hexagon and a unit cell formed by three sublattices \cite{wang2023quantum,1g9n-wm38}.  Although pristine kagome lattices, with  all corner-sharing triangles having the same size and bond length, were shown  to be relevant for extreme geometric frustration \cite{PhysRevLett.100.077203,PhysRevLett.104.147201,PhysRevLett.108.045305,PhysRevLett.131.256501}, distorting the kagome lattice opens a gap at the Dirac points that triggers topology  and    macroscopic charge density wave and superconducting correlations \cite{wang2023quantum,1g9n-wm38}. Interestingly, during a specific structural distortion, the triangles forming the kagome lattice alternately expand and contract, giving rise to a \emph{breathing mode} \cite{PhysRevLett.129.166401}, which turns out to be  a key mechanism for a plethora of emergent phases in kagome materials \cite{wang2023quantum,1g9n-wm38}.

An intriguing feature of breathing modes is that the structural distortion originating them breaks spatial inversion symmetry, making them a unique platform for designing inversion symmetry breaking phases. 
Examples involve, e. g., anomalous Hall states \cite{PhysRevB.102.035412},  2D multiferroics and magnetoelectric effect \cite{PhysRevB.104.L060405,PhysRevB.109.014433}, ferroelectricity-engineered valley \cite{zhao2024ferroelectrovalley}, spin filter transport \cite{xing2024independent}, flat band and many-body engineering \cite{regmi2022spectroscopic,wr7w-nfhg}, topological phases \cite{PhysRevB.105.085138,yjg9-4vmp,PhysRevB.105.085411,geschner2024,29ht-pwyt}, and superconductivity \cite{meena2024superconductivity,Liu23,Graham24}.  
Moreover, the breathing mode has been shown to be   highly  controllable, e. g., by means of electric fields \cite{29ht-pwyt} and by molecular orbital design in metal-organic frameworks \cite{lu2026tunable,yang2026breathing}, hence making kagome lattices with breathing modes a fertile platform  for emergent physics.

\begin{figure}[!t]
\centering
\includegraphics[width=0.48\textwidth]{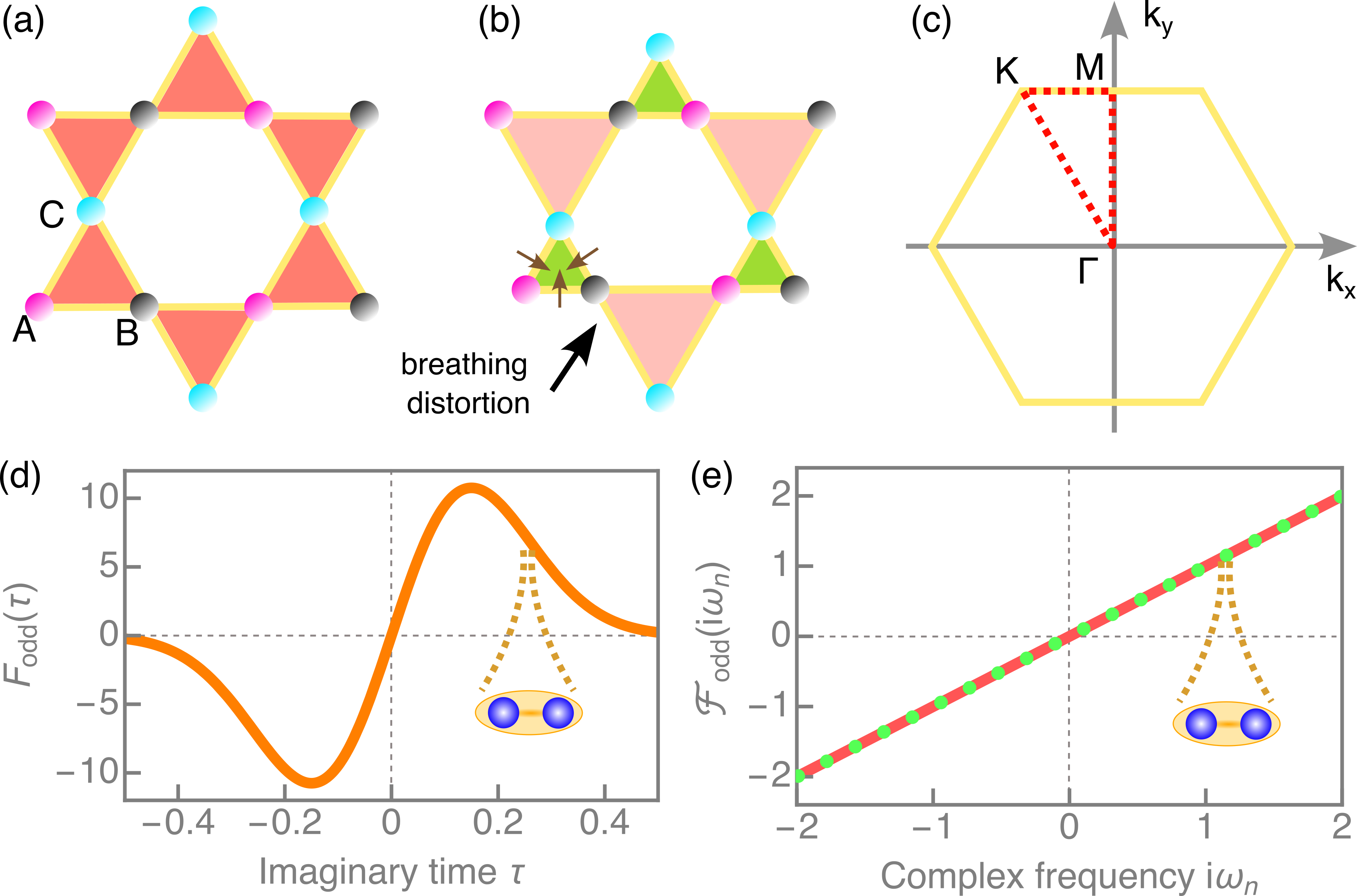}
\caption{(a) Sketch of a pristine regular kagome lattice (yellow)  made of corner-sharing triangles with sublattices A, B and C. (b) Kagome lattice under a breathing distortion (brown arrows), which originates the breathing mode. (c) First Brillouin zone of the kagome lattice indicating the high symmetry points and the path K$\Gamma$MK (red dashed lines). (d,e) Schematic pair amplitude induced by the breathing mode as a function of the imaginary time $\tau$ (d) and complex frequency in $i\omega_{n}$ (e). The resulting dynamical pair at finite times and frequencies is  depicted by the blueish filled circles inside the orange ellipse.}
\label{Fig1} 
\end{figure}

In this work, we demonstrate that the breathing mode in kagome superconductors induces dynamical odd-frequency pairing  that characterizes an emergent phase of Cooper pairs. To show this, we first perform a full symmetry classification of superconducting symmetries in kagome lattices [Fig.\,\ref{Fig1}(a-c)], which reveals that eight pair classes are naturally allowed, with four of them exhibiting odd-frequency symmetry. These odd-frequency Cooper pairs are entirely   nonlocal in time [Fig.\,\ref{Fig1}(d,e)] and hence of dynamical origin, while exhibiting singlet and triplet configurations that depend on the sublattices and parity. We then  prove that the emergent odd-frequency spin-singlet Cooper pairs can naturally emerge induced by the breathing mode in  kagome monolayers with conventional spin-singlet superconductivity. The prospect of inducing and controlling dynamical odd-frequency Cooper pairs by the breathing mode in kagome lattices opens the route for breathing engineered dynamical superconducting correlations.  

\textit{Allowed pair symmetries in kagome materials}.---We begin by inspecting all the possible types of Cooper pairs that are allowed to emerge in  kagome superconducting systems. For this purpose, given that  Cooper pairs are directly characterized by superconducting pair amplitudes \cite{zagoskin,mahan2013many,tanaka2011odd_review,Cayao2020odd,triola2020role}, we analyze   pair amplitude symmetries  taking into account the sublattice nature of kagome lattices under equilibrium and spatial translation conditions. The pair amplitude is   given by the anomalous Green's function 
\begin{equation}
\label{FF}
\mathcal{F}^{\alpha\alpha^{\prime}}_{\sigma\sigma'}(\bm{k},i\omega_n)=-\int_{0}^{\beta}d\tau\,{\rm e}^{i\omega_{n}\tau}\langle \mathcal{T}_{\tau}c^{\alpha}_{\bm{k}\sigma}(\tau)c^{\alpha^{\prime}}_{-\bm{k}\sigma'}(0)\rangle\,,
\end{equation}
where $\omega_n=(2n+1)\pi/\beta$ are fermionic Matsubara frequencies, with $\beta=1/(\kappa_{\rm B}T)$, while  $c^{\alpha}_{\bm{k}\sigma}(\tau)$ annihilates an electronic state with spin $\sigma$ and two-dimensional momentum $\bm{k}$ at imaginary time $\tau$ in the sublattice $\alpha=A,B,C$ since a kagome lattice consists of three sublattices \cite{wang2023quantum,1g9n-wm38}; here, $\mathcal{T}_{\tau}$ is the imaginary time ordering operator. To identify the type of allowed pair symmetry, we now account for the fermionic nature of electrons, which dictates an antisymmetry condition for the pair amplitude in Eq.\,(\ref{FF}) under the total exchange of quantum numbers, namely,  $\mathcal{F}^{\alpha\alpha^{\prime}}_{\sigma\sigma'}(\bm{k},z)=-\mathcal{F}^{\alpha^{\prime}\alpha}_{\sigma'\sigma}(-\bm{k},-z)$, with $z=i\omega_n$. This condition embodies  Fermi-Dirac statistics and enables a complete classification of  superconducting correlations allowed in kagome superconductors; similar ideas have been   shown to be useful in other superconducting systems \cite{RevModPhys.77.1321,tanaka2011odd_review,RevModPhys.91.045005,Cayao2020odd,triola2020role,tanaka2024review,FukayaJPCM2025}. Thus, under  the individual exchange of either frequency, spins, sublattices, or momentum, which are represented by involution operators with eigenvalues being $\pm1$, the pair amplitude becomes either \emph{even} (E) or \emph{odd} (O); note that odd and  even under spin exchange defines a spin-singlet (S) and spin-triplet (T) symmetry, respectively. With these binary possibilities for each exchange, there is $2^3=8$ possibilities where the pair amplitude is fully antisymmetric and, therefore, consistent with  Fermi-Dirac statistics. These eight pair symmetry classes are shown in Table \ref{Table1} and represent all the types of Cooper pairs allowed in kagome superconductors.

\begin{table}[t!]
\centering
\begin{tabular}{ |c|c|c|c|c|  }
 \hline
 {\bf Frequency}& {\bf  Spin}& {\bf Sublattice}& {\bf Momentum}&{\bf Class} \\[0.5ex] 
 $z\rightarrow-z$ & $\sigma\rightarrow\sigma'$ &  $\alpha \rightarrow\beta$&  $\boldsymbol{k}\rightarrow-\boldsymbol{k}$ & total \\
  \hline
  \hline
  Even&Triplet&Even&Odd& ETEO
\\
  \hline
  Even &Triplet&Odd&Even&ETOE\\
  \hline
 Even &Singlet&Even&Even&ESEE\\
  \hline
  Even &Singlet&Odd&Odd&ESOO\\
 \hline
  Odd & Triplet&Even&Even&OTEE
\\
  \hline
  Odd &Triplet&Odd&Odd&OTOO\\
  \hline
  Odd &Singlet&Even&Odd&OSEO\\
  \hline
  Odd &Singlet&Odd&Even&OSOE\\
 \hline
\end{tabular}
\caption{All possible symmetry classes for  pair amplitudes allowed in kagome superconductors.}
\label{Table1}
\end{table}

\begin{figure}[!t]
\includegraphics[width=1.0\linewidth]{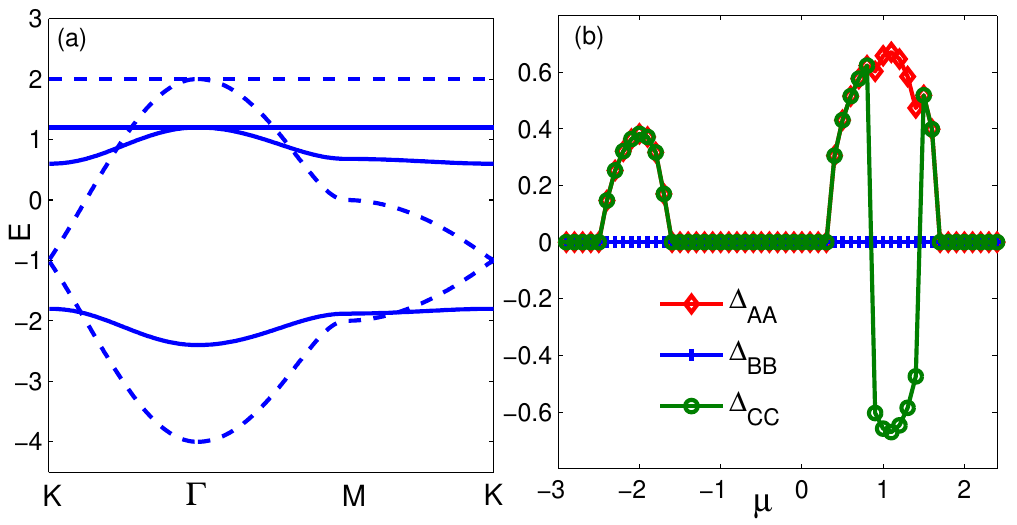} \caption{(a) Energy bands of the normal state Hamiltonian Eq.~\eqref{eq:hamil2} for $t_b=t_a$ (dashes lines) and $t_b=0.2t_a$ (solid lines) with $\mu=0$. (b) Self-consistent values of the superconducting order parameters $\Delta_{AA,BB,CC}$ with varying $\mu$ for $t_b=0.2t_a$.}
\label{fig:bandsanddelta} 
\end{figure}

\textit{Modelling superconducting kagome materials}.---In order to investigate the emergence of dynamical superconducting pairings among the ones mentioned in Table \ref{Table1}, we need to first consider and solve the parent superconducting Hamiltonian with breathing distortion [Fig.\,\ref{Fig1}(a,b)]. The breathing kagome lattice can be modeled using different nearest neighbor hoppings $t_a$ and $t_b$. The corresponding Hamiltonian in the three sublattice basis $\Psi_{\bm{k},\sigma}=\left(c^{A}_{\bm{k}\sigma},c^{B}_{\bm{k}\sigma},c^{C}_{\bm{k}\sigma}\right)^{\rm T}$ can be written as $\mathcal{H}_n=\sum_{\bm{k},\sigma} \Psi^{\dagger}_{\bm{k},\sigma} \hat{\mathcal{H}}_n(\bm{k}) \Psi_{\bm{k},\sigma}$ with
\begin{equation}
\hat{\mathcal{H}}_n(\bm{k})=\left(\begin{array}{ccc} -\mu & t_a+t_be^{-i\bm{k}\cdot\bm{a}_3} & t_a+t_be^{-i\bm{k}\cdot\bm{a}_2}  \\
t_a+t_be^{i\bm{k}\cdot\bm{a}_3} & -\mu  & t_a+t_be^{-i\bm{k}\cdot\bm{a}_1}  \\
t_a+t_be^{i\bm{k}\cdot\bm{a}_2} & t_a+t_be^{i\bm{k}\cdot\bm{a}_1} & -\mu  \\
\end{array}\right),
\label{eq:hamil2}
\end{equation}
where $\mu$ is the chemical potential, $\bm{a}_{1,2}=(\pm1/2,\sqrt{3}/2)$ and $\bm{a_3}=\bm{a_2}-\bm{a_1}$. In Eq.~\eqref{eq:hamil2}, $t_a=t_b$ represents a perfect kagome lattice with no breathing distortion, while $t_a\neq t_b$ induces a breathing distortion that characterizes the breathing mode. In Fig.~\ref{fig:bandsanddelta}(a), we show the effect of having breathing distortion on the band structure along a high-symmetry path [Fig.\,\ref{Fig1}(c)]. Typical of a kagome lattice, $t_a=t_b=-t$ ($t=1$ is taken as the energy unit), shown by dashed curves in Fig.~\ref{fig:bandsanddelta}(a) at $\mu=0$, features  Dirac points at $K$, van Hove singularities at the $M$ point, and a flat band. In contrast, the breathing distortion introduces a gap at $K$ due to broken inversion symmetry, as shown by solid lines in Fig.~\ref{fig:bandsanddelta}(a) for $t_b=0.2t_a$, $t_a=-t$, and $\mu=0$.

To assess superconductivity, we consider the simplest onsite spin-singlet superconductivity on the breathing kagome lattice. To keep the discussion generic, we consider three different superconducting order parameters $\Delta_{AA}$, $\Delta_{BB}$, $\Delta_{CC}$ for the three different sublattices A,B,C, respectively. The order parameters are obtained using the self-consistent conditions
\begin{equation}
\Delta_{\alpha\alpha}=-V\sum_{\bm{k}^{\prime}}\langle c^{\alpha}_{\bm{k}^{\prime} \uparrow}c^{\alpha}_{-\bm{k}^{\prime}\downarrow} \rangle,
\label{eq:selfcon}
\end{equation}
where $V$ is the attractive interaction strength. Due to breathing distortion, $V$ can be different for different sublattices. The total superconducting Hamiltonian can be written in the Bogoliubouv de-Gennes form in the Nambu basis $\Psi^{\dagger}_{\rm{BdG}}=\left(c_{\bm{k}\uparrow}^{A\dagger},c_{\bm{k}\uparrow}^{B\dagger},c_{\bm{k}\uparrow}^{C\dagger},c^{A}_{-\bm{k}\downarrow},c^{B}_{-\bm{k}\downarrow},c^{C}_{-\bm{k}\downarrow}\right)$ as $\mathcal{H}_{\rm BdG}=\sum_{\bm{k}}\Psi^{\dagger}_{\rm{BdG}} \hat{\mathcal{H}}_{\rm BdG}(\bm{k})\Psi_{\rm{BdG}}$ with
\begin{equation}
\hat{\mathcal{H}}_{\rm BdG}=\left(\begin{array}{cc} \hat{\mathcal{H}}_n(\bm{k}) & \hat{\mathcal{H}}_{\Delta}  \\
\hat{\mathcal{H}}^{\dagger}_{\Delta} & -\hat{\mathcal{H}}^{*}_n(-\bm{k})   \\
\end{array}\right),
\label{eq:bdgmatrix}
\end{equation}
where
\begin{equation}
\hat{\mathcal{H}}_{\Delta}=\left(\begin{array}{ccc} \Delta_{AA} & 0 & 0  \\
0 & \Delta_{BB} & 0  \\
0 & 0 & \Delta_{CC}  \\
\end{array}\right).
\label{eq:deltamatrix}
\end{equation}
We diagonalize the Hamiltonian $\hat{\mathcal{H}}_{\rm BdG}$ and solve the self-consistent equations of $\Delta_{AA,BB,CC}$ iteratively using Eq.~\eqref{eq:selfcon}. We use a system size of 25950 sites and $V=2.0$ as the interaction strength.

Onsite superconductivity can give three different irreducible representations for the momentum dependence \cite{Holbaek23}, namely $A_1$, $E_2^{(1)}$, and $E_2^{(2)}$. For $A_1$ representation, $\Delta_{AA}=\Delta_{BB}=\Delta_{CC}$. We have verified that this solution is possible both with and without breathing distortion. Other symmetries $E_2^{(1)}$ and $E_2^{(2)}$ require $\Delta_{CC}=-2\Delta_{BB}=-2\Delta_{CC}$ and $\Delta_{AA}=-\Delta_{CC}$, $\Delta_{BB}=0$, respectively. We do not find $E_2^{(1)}$ for any doping with both $t_a=t_b$ and $t_a \neq t_b$ even with a bias towards such a solution. Interestingly, when we bias the solution towards $E_2^{(2)}$, we find that such a solution is absent for $t_a=t_b$, but present when $t_a \neq t_b$. We show this solution in Fig.~\ref{fig:bandsanddelta}(b) for $t_b=0.2t_a$. Superconductivity is finite for $-2.4<\mu<-1.7$ and $0.4<\mu<1.6$ which correspond to metallic normal state, see the bands in Fig.~\ref{fig:bandsanddelta}(a). Interestingly, for $0.9<\mu<1.4$ we get a pure $E_2^{(2)}$ where $\Delta_{BB}=0$ and $\Delta_{AA}=-\Delta_{CC}$. This is the same energy regime where the dispersive middle band touches the flat band near $\Gamma$ point. The appearance of $E_2^{(2)}$ near the band touching point is related to the sublattice texture of the bands. Near $\Gamma$ point, the middle band has pure A-sublattice while the flat band has pure C-sublattice \cite{Jalalimola23}. Although such sublattice texture is also present for $t_b=t_a$, the $E_2^{(2)}$ does not appear for $t_b=t_a$ due to a significant dispersion of the middle band. In contrast, for $t_b \neq t_a$ the middle band becomes flatter with enhanced density of states favoring $E_2^{(2)}$ states. A superconducting state with unequal intra-sublattice pairing with $E_2^{(2)}$ configuration breaks the $C_3$ symmetry of the kagome lattice as it singles out one of the sublattices.

\textit{Emergent odd-frequency dynamical pairing}.---After obtaining the self-consistent solutions of $\hat{\mathcal{H}}_{\rm BdG}$ in Eq.~\eqref{eq:bdgmatrix}, the superconducting pair amplitudes are obtained from the anomalous part of the total Nambu Green's function $[z-\hat{\mathcal{H}}_{\rm BdG}(\bm{k})]\mathcal{G}(z,\bm{k})=\mathbb{I}$. Although under general circumstances there appear intra- and inter-sublattice superconducting correlations, the former are particularly intriguing because they entirely depend on the structural distortion of the kagome lattice known as breathing mode, as we demonstrate next. For the intra-sublattice superconducting pair amplitudes, we find
\begin{equation}
\label{FAA}
\mathcal{F}_{\uparrow\downarrow,{\rm O}}^{\rm AA}(\bm{k},z)=\frac{4i\,z\,t_{a}t_{b}(t_{a}-t_{b})(\Delta_{\rm BB}-\Delta_{\rm CC})\mathcal{S}_{\bm{k}}}{\mathcal{D}(\bm{k},z)}\,,
\end{equation}
where $\mathcal{S}_{\bm{k}}=[{\rm cos}(k_{x}/2)-{\rm cos}(\sqrt{3}k_{y}/2)]{\rm sin}(k_{x}/2)$ and $\mathcal{D}(\bm{k},z)$ is a sixth degree polynomial in $z$ and an  even function of $z$. We find that $\mathcal{F}_{\uparrow\downarrow,{\rm O}}^{\rm BB}$ and $\mathcal{F}_{\uparrow\downarrow,{\rm O}}^{\rm CC}$ can be obtained by performing the following replacements $({{\rm B}\rightarrow {\rm A}},{{\rm C}\rightarrow {\rm C}})$ and $({{\rm B}\rightarrow {\rm A}},{{\rm C}\rightarrow {\rm B}})$, respectively. By a close inspection of the emergent superconducting pair given by Eq.\,(\ref{FAA}), we identify that it has odd-momentum parity due to the ${\rm sin}(k_x/2)$ through $\mathcal{S}_{\bm{k}}$ and develops an odd-frequency dependence via the complex frequency $z$ in the numerator. Therefore, $\mathcal{F}_{\uparrow\downarrow,{\rm O}}^{\rm \alpha\alpha}(\bm{k},z)$ ($\alpha=A,B,C$) exhibits an odd-frequency, spin-singlet, even-sublattice, odd-momentum parity symmetry, which belongs to the OSEO class in Table \ref{Table1}. Interestingly, as evident from $\mathcal{S}_{\bm{k}}$, $\mathcal{F}_{\uparrow\downarrow,{\rm O}}^{\rm \alpha\alpha}(\bm{k},z)$ has a $B_1$ symmetry, which typically arises from nearest neighbor pairing potential driven by electronic repulsion \cite{Romer22}, here appears only due to onsite pairing potential in the Hamiltonian Eq.~\eqref{eq:deltamatrix}. Notably, due to the odd-frequency nature, $\mathcal{F}_{\uparrow\downarrow,{\rm O}}^{\rm \alpha\alpha}(\bm{k},z)$ is spin-singlet in nature in contrast to the triplet nature considered in Ref.~\onlinecite{Romer22}.

To realize $\mathcal{F}_{\uparrow\downarrow,{\rm O}}^{\rm AA}(\bm{k},z)\neq0$, it requires the numerator of the expression given by Eq.\,(\ref{FAA}) to be nonzero, which is achieved within the Brillouin zone when $\Delta_{\rm BB}\neq \Delta_{CC}$ and $t_{a}\neq t_{b}\neq0$ at finite frequencies. Notably, since distinct sublattices are very likely to have  distinct pair potentials as demonstrated in Fig.\,\ref{fig:bandsanddelta}(b), the nonzero value of the odd-frequency pair amplitude $\mathcal{F}_{\uparrow\downarrow,{\rm O}}^{\rm \alpha\alpha}(\bm{k},z)$ entirely relies on the second condition ($t_{a}\neq t_{b}\neq0$), which encodes the structural distortion due to the breathing mode. Therefore, the breathing mode induces Cooper pairs having an odd-frequency, spin-singlet, even-sublattice, odd-momentum parity symmetry as an emergent phenomenon in kagome materials with conventional parent spin-singlet $s$-wave superconductivity. Intriguingly, the odd-frequency nature of OSEO pairs makes them nonlocal in the relative time between paired electrons [Eq.\,(\ref{FF})], hence signaling an entirely dynamical phenomenon arising due to the breathing mode. Besides  odd-frequency Cooper pairs \footnote{Furthermore, we also obtain finite inter-sublattice pair amplitudes, which develop OSEO and ESEE classes but also OSOE symmetry, all consistent with that allowed pair symmetries in kagome superconductors given in Table \ref{Table1}.}, we also find even-frequency pairs with spin-singlet, even-parity symmetry, denoted by $\mathcal{F}_{\uparrow\downarrow,{\rm E}}^{\rm AA}(\bm{k},z)$, and correspond to the ESEE class in Table \ref{Table1}. We note that these ESEE Copper pairs originate  from the spin-singlet $s$-wave parent superconductor but also get a contribution due to the breathing mode. Thus, while ESEE Cooper pairs exist even without the breathing mode, OSEO pairs are entirely tied to the presence of the breathing mode.

\begin{figure}[!t]
\centering
\includegraphics[width=0.49\textwidth]{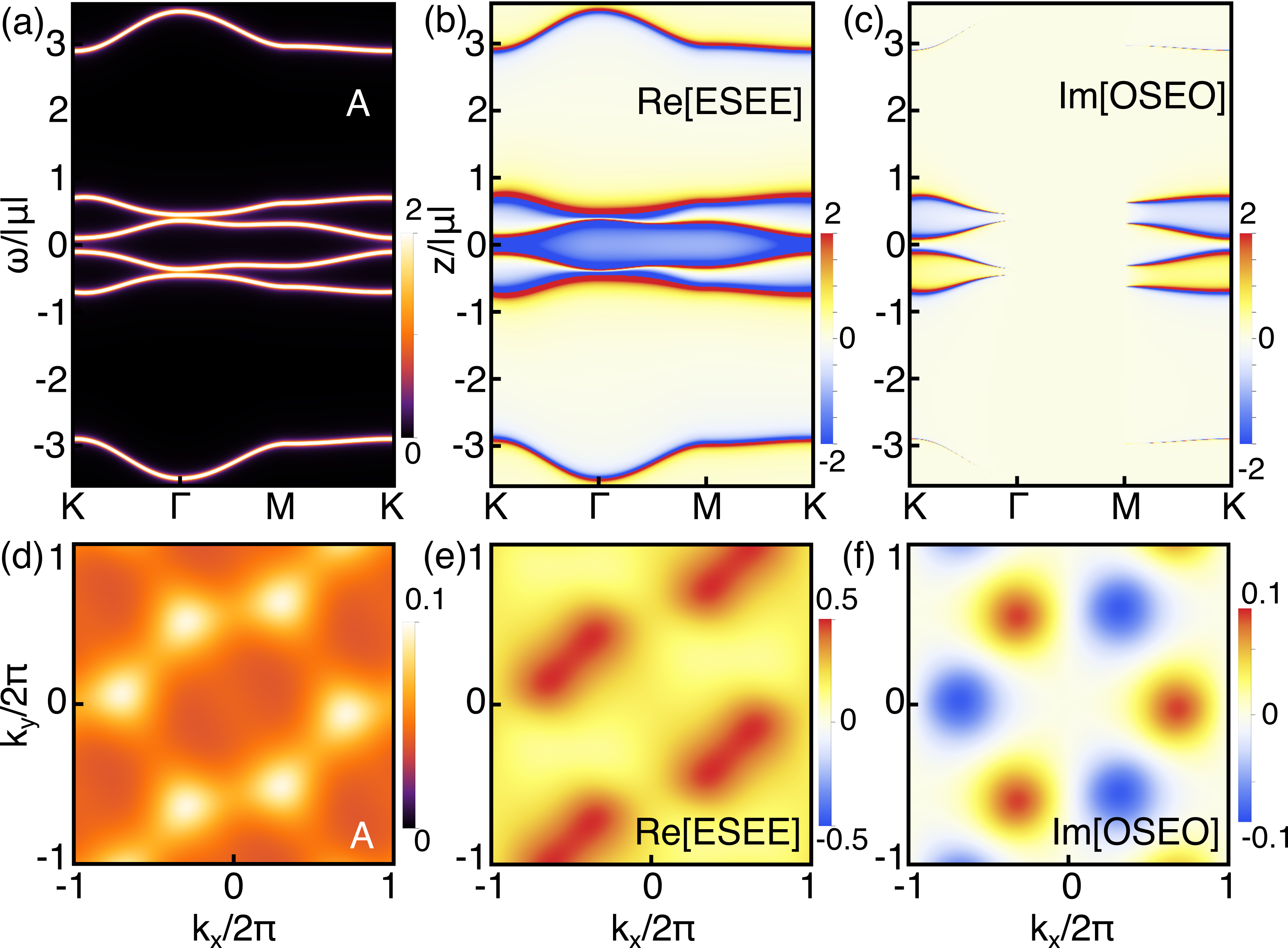}
\caption{(a-c) Spectral function, ESEE and OSEO pair amplitudes as functions of frequency along high-symmetry paths under a finite breathing distortion $t_{a}\neq t_{b}$. The spectral function is plotted versus $\omega$, with $z\rightarrow \omega+i\eta$, while the pair amplitudes is shown for $z$.    (d-f) The same as in (a-c) but as a function of momenta at $\omega=1$ for the spectral function, while $z=1$ for the pair amplitudes. Parameters: $\Delta_{\rm AA}=0.66|\mu|$, $\Delta_{\rm BB}=0$, $\Delta_{\rm CC}=-0.66|\mu|$, $\mu=1$, $t_{b}=0.2t_{a}$ when not specified.}
\label{Fig3} 
\end{figure}

To visualize the emergence of dynamical OSEO pairs, in Fig.~\ref{Fig3} we present its pair amplitude, in particular $\mathcal{F}_{\uparrow\downarrow,{\rm O}}^{\rm BB}(\bm{k},z)$, along with the ESEE counterpart and the spectral function $\mathcal{A}(\bm{k},\omega)$ under a finite breathing distortion $t_{a}\neq t_{b}$. In Fig.~\ref{Fig3}, we focus on one parameter point in Fig.~\ref{fig:bandsanddelta}, $\mu=1$, and take the self-consistent values of the $E_2^{(2)}$ pairing obtained earlier. Since $\mu=1$ lies close to the band touching point of the dispersive and flat bands of the normal state, the BdG spectrum shown in (a) by plotting $\mathcal{A}(\bm{k},\omega)$ has a hybridization of these two bands. Hence, the normal state flat band becomes dispersive in the superconducting spectrum with the sublattice textures of the flat band getting reflected in the emergent dynamical pairings. ESEE attains high values near BdG bands, see the similarity of (a) and (b). For a fixed $z$, ESEE does not show any nodes or sign change in the pairing amplitude, as shown in (e). In contrast, OSEO acquires a distinct odd-parity symmetry with the form factor $\mathcal{S}_{\bm{k}}$. The nodes of the OSEO component are apparent in (c) between $\Gamma$ and $M$ where the path is along $k_x=0$ where $\mathcal{S}_{\bm{k}}=0$. The oddness in frequency is also evident from the sign reversal in OSES between $z<0$ and $z>0$. For a fixed $z$, OSEO shown for the full Brillouin zone in (f) also shows the odd-parity nature reflected by the sign-reversal among the lobes between $k_x<0$ and $k_x>0$. Moreover, we see that the $C_3$ symmetry breaking in the $E_2^{(2)}$ is reflected as a rotational asymmetry seen in (d,e,f). Although in Fig.~\ref{Fig3} we show only $\mathcal{F}_{\uparrow\downarrow,{\rm O}}^{\rm BB}(\bm{k},z)$, other OSEO pairings $\mathcal{F}_{\uparrow\downarrow,{\rm O}}^{\rm AA/CC}(\bm{k},z)$ are also present and behave similarly. Also note that the momentum dependence of ESEE and OSEO shown in Fig.~\ref{Fig3} is similar for the whole range $0.9<\mu<1.4$ where $E_2^{(2)}$ is obtained. For other $\mu$ values where we find superconductivity in Fig.~\ref{fig:bandsanddelta}, $\mathcal{F}_{\uparrow\downarrow,{\rm O}}^{\rm AA/CC}(\bm{k},z)\neq0$ but $\mathcal{F}_{\uparrow\downarrow,{\rm O}}^{\rm BB}(\bm{k},z)=0$ since $\Delta_{BB}=0$ and $\Delta_{AA}=\Delta_{CC}$.

\textit{Detection of breathing-induced odd-frequency pairs}.--- Odd frequency dynamical pairing can be detected experimentally using various techniques. These include Andreev conductance \cite{KT2000,PhysRevB.54.7366,bwp9-7dsd,BOOKASANO},  Doppler shift measurements as those predicted in Refs.\,\cite{TanakaJPSJ2002,TanumaPRB2002_Sep,TanumaPRB2002_Nov,Tanaka2003JPSJ}, ARPES \cite{RevModPhys.75.473,RevModPhys.93.025006,zhang2022angle,RevModPhys.96.015003}, fluctuation spectroscopy \cite{Kornich21}, paramagnetic Meissner effect \cite{PhysRevLett.125.026802,PhysRevX.5.041021,PhysRevB.101.180512} and quasiparticle interference experiments~\cite{allan2012anisotropic,allan2013imaging,marques2021tomographic,wang2025NatPhys,Perrin20,Chakraborty22a}. In particular, quasiparticle interference is a widely used method to determine pairing symmetries in superconductors \cite{Hanaguri09,Liu19,Cheung20,Boker20,Hoffman02a, Wang03, McElroy03,Sharma21}, and was also proposed for kagome superconductors \cite{Liu24}. Quasiparticle interference probes the change in the local density of states in the presence of impurities. The change in the local density of states can be written using the Nambu Green's function in momentum space $\mathcal{G}(\omega,\bm{k})$ as \cite{Hirschfeld15,Altenfeld18,Liu19,Cheung20,Boker20,Sharma21},
\begin{equation}
\delta\rho_{\alpha}(q,\omega)=-\frac{1}{\pi}{\rm Im}\left[\sum_{k}\mathcal{G}(\omega,\bm{k})T\mathcal{G}(\omega,\bm{k+q})\right]_{\alpha\alpha},
\label{eq:rhoqdef}
\end{equation}
where $T$ is the T-matrix \cite{MahanBook} corresponding to the impurity. For weak non-magnetic impurities, $T=V_{\text {imp}}\sigma_0 \bigotimes\tau_3$ \cite{Hirschfeld15}, where $\tau_3$ and $\sigma_0$ are in the Nambu and sublattice bases, respectively, and $V_{\text {imp}}$ is the impurity strength. Eq.~\eqref{eq:rhoqdef} connects the odd-frequency dynamical pairing [Eq.\,(\ref{FAA})] to the observable $\delta\rho_{\alpha}(q,\omega)$, hence offering a direct detection of    odd-frequency Cooper pairs through quasiparticle interference experiments \cite{Chakraborty22a} in  breathing kagome superconducting lattices. Currently, there already exist compounds harboring breathing kagome lattices and superconductivity, such as Ta$_{2}$V$_{3.1}$Si$_{0.9}$ with a relatively high T$_c$=7.5K \cite{Liu23}, which places  breathing-induced odd-frequency Cooper pairs within experimental reach.

\textit{Conclusions and discussion}.---
In conclusion, we have demonstrated that kagome lattices harbor a plethora of Cooper pair types, with the most intriguing family being those possessing a dynamical odd-frequency pairing due to their intrinsic nonlocal in time nature. In particular, we have found that  odd-frequency $p$-wave Cooper pairs can be entirely driven by the breathing distortion in kagome materials with conventional spin-singlet $s$-wave  superconductivity. Although  breathing-induced odd-frequency Cooper pairs require a sublattice symmetry breaking in superconducting order parameters, primarily considered here by the $E_2^{(2)}$ order, we stress that it is an unavoidable situation and, therefore, expected to occur due to other various reasons. For example, the interaction strengths driven by breathing phonons can be different for different sublattices. Another ubiquitous possibility is the presence of impurities; impurities reside in real space sites and are random. Hence, impurity strengths of different sublattices are definitely different for random disorder, and consequently, the superconducting order parameters for different sublattices will be different. Yet another way can be by proximitizing a breathing kagome lattice to a superconductor, where superconductivity is induced by proximity effect and distinct pair potential would necessarily lead to dynamical odd-frequency Cooper pairs driven by the breathing mode.  Besides onsite parent superconductivity,  we expect that our symmetry analysis is also applicable to kagome lattices with nearest neighbor superconductivity \cite{Wu21,Romer22,Wu24} or even under the presence of charge density waves \cite{Chakraborty22,Chakraborty21}. Our findings, therefore, establish breathing kagome lattices as a rich playground for realizing dynamical odd-frequency Cooper pairs.

A. D. acknowledges financial support from Keele Foundation and Magnus Ehrnrooth Foundation as part of the SuperC collaboration. J. C. acknowledges financial support from  the Swedish Research Council (Vetenskapsr{\aa}det Grant No. 2021-04121), and from the Olle Engkvist Foundation (Grant No.  243-1026). The computations were enabled by resources provided by the National Academic Infrastructure for Supercomputing in Sweden (NAISS), partially funded by the Swedish Research Council through Grant Agreement No. 2022-06725.
  
\bibliography{biblio}

\end{document}